\documentclass[aj,twocolumn]{aastex62}

\usepackage{placeins}
\usepackage{comment}
\usepackage{graphicx}
\newcommand{\thisstar}{HIP 94235}

\usepackage{url}
\usepackage{hyperref}

\usepackage{wrapfig}

\usepackage{lineno}

\submitjournal{AJ}

\shorttitle{Ly-$\alpha$ Observations of a 120 Myr Old Sub-Neptune}
\shortauthors{Morrissey et al.}

\begin{document}

\title{Searching for Neutral Hydrogen Escape from the 120 Myr Old Sub-Neptune HIP94235b using HST}

\correspondingauthor{Ava Morrissey}
\email{ava.morrissey@unisq.edu.au}

\author[0009-0000-3527-8860]{Ava Morrissey} 
\affil{University of Southern Queensland, Centre for Astrophysics, West Street, Toowoomba, QLD 4350, Australia}

\author[0000-0002-4891-3517]{George Zhou}
\affil{University of Southern Queensland, Centre for Astrophysics, West Street, Toowoomba, QLD 4350, Australia}

\author[0000-0003-0918-7484]{Chelsea X. Huang}
\affil{University of Southern Queensland, Centre for Astrophysics, West Street, Toowoomba, QLD 4350, Australia}

\author[0000-0001-7294-5386]{Duncan Wright}
\affil{University of Southern Queensland, Centre for Astrophysics, West Street, Toowoomba, QLD 4350, Australia}

\author[0009-0004-0425-0813]{Caitlin Auger}
\affil{University of Southern Queensland, Centre for Astrophysics, West Street, Toowoomba, QLD 4350, Australia}

\author[0000-0003-1337-723X]{Keighley E. Rockcliffe}
\affil{Department of Physics and Astronomy, Dartmouth College, Hanover, NH 03755, USA}

\author[0000-0003-4150-841X]{Elisabeth R. Newton}
\affil{Department of Physics and Astronomy, Dartmouth College, Hanover, NH 03755, USA}

\author[0000-0001-7615-6798]{James G. Rogers}
\affil{Department of Earth, Planetary, and Space Sciences, University of California, Los Angeles, 595 Charles E. Young Drive East, Los Angeles, CA 90095, USA}

\author[0000-0002-9308-2353]{Neale Gibson}
\affil{School of Physics, Trinity College Dublin, University of Dublin, Dublin 2, Ireland}

\author[0000-0001-6508-5736]{Nataliea Lowson}
\affil{University of Southern Queensland, Centre for Astrophysics, West Street, Toowoomba, QLD 4350, Australia}

\author[0000-0002-4321-4581]{L. C. Mayorga}
\affil{The Johns Hopkins University Applied Physics Laboratory, 11100 Johns Hopkins Road, Laurel, MD 20723, USA}

\author[0000-0001-9957-9304]{Robert A. Wittenmyer}
\affil{University of Southern Queensland, Centre for Astrophysics, West Street, Toowoomba, QLD 4350, Australia}



\begin{abstract}
    HIP94235 b, a 120 Myr old sub-Neptune, provides us the unique opportunity to study mass loss at a pivotal stage of the system's evolution: the end of a 100 million year (Myr) old phase of intense XUV irradiation. We present two observations of HIP94235 b using the Hubble Space Telescope's Space Telescope Imaging Spectrograph (\textit{HST}/STIS) in the Ly-$\alpha$ wavelength region. We do not observe discernible differences across either the blue and red wings of the Ly-$\alpha$ line profile in and out of transit, and report no significant detection of outflowing neutral hydrogen around the planet. We constrain the rate of neutral hydrogen escaping HIP94235 b to an upper limit of $10^{13}$ gs$^{-1}$, which remains consistent with energy-limited model predictions of $10^{11}$ gs$^{-1}$. The Ly-$\alpha$ non-detection is likely due to the extremely short photoionization timescale of the neutral hydrogen escaping the planet's atmosphere. This timescale, approximately 15 minutes, is significantly shorter than that of any other planets with STIS observations. Through energy-limited mass loss models, we anticipate that HIP94235 b will transition into a super-Earth within a timescale of 1 Gyr. 
\end{abstract}



\section{Introduction}
\label{sec:introduction}
Lyman-$\alpha$ (Ly-$\alpha$) transits are invaluable tools for probing atmospheric escape from exoplanets, particularly those on close-in orbits to their host stars. These transits, which allow the tracing of neutral hydrogen outflows, provide vital insights into ongoing mass-loss processes and planet-star interactions.

\textit{HST} observations of Gyr old Neptunes have revealed extended hydrogen exospheres and tails about GJ 436b \citep{Kulow_2014,2015Natur.522..459E,Lavie_2017}, GJ 3470b \citep{2018AA...620A.147B}, and HAT-P-11 b \citep{2022NatAs...6..141B}, with a tentative detection for K2-18b \citep{2020A&A...634L...4D}. Despite their importance across all evolutionary stages, observations of neutral hydrogen outflows have traditionally been focused on older ($\sim$ Gyr old) planets, leaving a gap in our understanding of early atmospheric evolution. The mass-loss rates inferred from these older planets are consistent with a later stage evolution of their atmosphere, and do not test the rapid evolution expected of planets early on in their lives.

For small planets hosting gaseous hydrogen-helium envelopes, we expect the mass-loss rate to be much more rapid in the first hundred million years as the planets undergo runaway hydrodynamic evaporation \citep{Owen_2017,2018MNRAS.476..759G}. Observing active mass-loss in newly born planets helps test this model. Recent planet discoveries from \emph{TESS} have unveiled a number of young systems at the $\sim 100$ Myr age range \citep[e.g.][]{2019ApJ...880L..17N,2020AJ....160...33R,2020Natur.582..497P,2021AJ....161...65N,Mann_2022,Wood_2023}, but few are close enough for Ly-$\alpha$ transit observations due to the absorbing effects of the interstellar medium (ISM). Exceptions to this include the 22 Myr old multi-planet system AU Mic (\citealt{Rockcliffe_2023}; Rockcliffe, in prep), and the 400 Myr old sub-Neptune multi-planet system HD63433 \citep{Zhang_2022}, both of which have been subjects of Ly-$\alpha$ observation studies, revealing evidence of ongoing atmospheric outflows. These observations confirm that AU Mic b and HD 63433 c are likely born with large H/He envelopes. Similar inferences have been made for V1298 Tau b, where WFC3/G141 transmission spectra \citep{Barat_2023} found the planet hosts a low density atmosphere, and from which an envelope mass fraction $M_{frac,env}$ $\sim$ 0.4 was inferred.

HIP94235 b is a 120 Myr old sub-Neptune orbiting its G-type host ($T_{eff}$ = 5991 K, 1.094 $M_\odot$, 1.08 $R_\odot$) on a 7.7 day period \citep{Zhou_2022}. HIP94235 b's age is critical as it allows for a direct investigation into the effects of photoevaporation, which is a dominant mass-loss mechanism at time scales of 100 Myr. HIP94235 b is smaller (R = $3.00^{+0.32}_{-0.28}$$R_\oplus$) than other young and pre-main sequence Neptunes, e.g. AU Mic b (R = $3.96\pm0.15$$R_\oplus$, \citealt{Wittrock_2023}), and as such may be more representative of the smaller super-Earth class of exoplanets that we see around mature-aged stars. The age and size of HIP94235 b offer a unique opportunity to study the transition from a sub-Neptune to a super-Earth over planetary timescales.

We present two \textit{HST}/STIS FUV-MAMA transit observations (\textit{HST}-GO-17152; PI: Zhou) of HIP94235 in Section~\ref{sec:obs}, including the spectral extraction and lightcurve analysis. We model the stellar Ly-$\alpha$ line, photoionization rate and mass-loss rate in Section~\ref{sec:discussion}, where we compare observed and expected mass-loss rates for HIP94235 b.

\section{Ly-$\alpha$ Observations and analysis}
\label{sec:obs}
\subsection{STIS observations}
We obtained two \textit{HST} transits of HIP94235 b with the STIS UV channel and the G140M grating (found in MAST:\dataset[10.17909/r7j2-8287]{http://dx.doi.org/10.17909/r7j2-8287}). Observations were centered at 1222\AA{} and employed a $52\times0.1$" slit configuration to achieve a spectral resolution of $R\sim10,000$ over the wavelength regime of 1194-1249 \AA{}. Each transit visit included five consecutive orbits, as longer visits are generally prohibited by South Atlantic Anomaly (SAA) crossings. The transit duration of HIP94235 b in white light is 2.4 hours, covering two \textit{HST} orbits. Both visits, on 2023-02-18 and 2023-08-14, were timed such that orbit 1 occurred pre-ingress, orbits 2 and 3 occurred in-transit, and orbits 4 and 5 occurred post-egress. Each orbit had a usable visibility of 50 minutes. The first orbit of each visit included an acquisition and acquisition/peak sequence, a 1528 s science integration, and a wavelength calibration exposure after the end of target visibility. Subsequent orbits consisted of a 2536 s science integration. Every orbit was taken in time-tag mode where each photon impact time on the detector is recorded and allows for finer time-resolved flux behavior to be analyzed.

\subsection{Spectral extraction}
Reduction and spectral extraction were performed with the official \textsc{stistools}\footnote{Latest version of \textsc{stistools} available at: \href{https://stistools.readthedocs.io/en/latest/}{https://stistools.readthedocs.io/en/latest/}} package, utilising the \textsc{calstis} pipeline to perform calibrations and spectral extractions. Each time-tagged exposure was broken down in to four sub-exposures with \textsc{inttag} in order to gain greater temporal resolution. The initial sub-orbit was 382 s and the remaining sub-orbits were 634 s.

The \textsc{calstis} pipeline executed various calibrations on the raw files, including bias and plane subtraction via the \textsc{BIASCORR} and \textsc{BLEVCORR} functions to remove constant patterns across the electronic zero-point of each CCD readout. Cosmic ray mitigation was performed using \textsc{CRCORR}, followed by flat fielding with \textsc{FLATCORR}.

We noted the presence of an extensive and varying background in the calibrated frames. This large background pattern varied at the orbit level, and can be attributed to the FUV-MAMA dark current\footnote{More information on STIS's dark current can be found at: \href{https://hst-docs.stsci.edu/stisdhb/chapter-4-stis-error-sources/4-1-error-sources-associated-with-pipeline-calibration-steps}{https://hst-docs.stsci.edu/stisdhb/chapter-4-stis-error-sources/4-1-error-sources-associated-with-pipeline-calibration-steps}}, which is intrinsic to the micro-channel plated array rather than phosphorescence. This intermittent glow is dependant on both the detector temperature and the duration the detector has been operational. Our observations commenced directly after \textit{HST} resumed operation following its daily shut down during its passage through the SAA, which explains the steady increase in dark current per orbit.

\begin{figure*}
    \centering
    \includegraphics[width=0.8\linewidth,trim={0 1.5cm 0 0},clip]{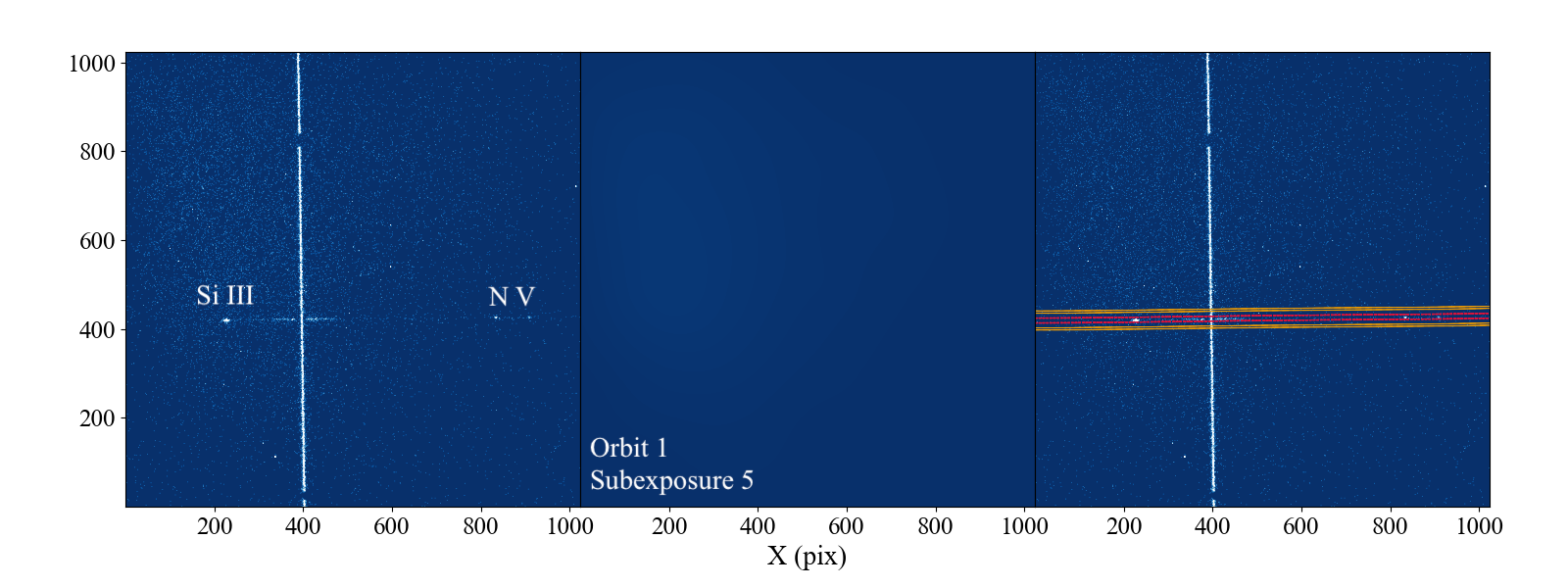}
    \includegraphics[width=0.8\linewidth,trim={0 1.5cm 0 1.5cm},clip]{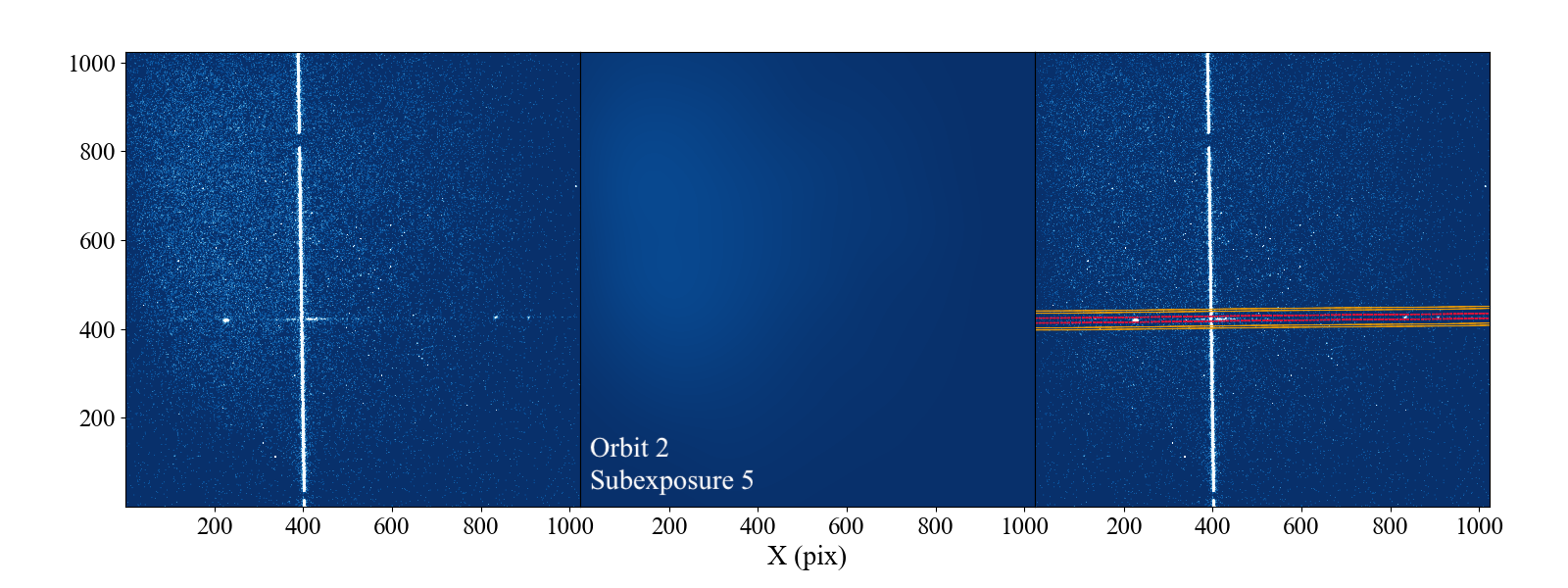}
    \includegraphics[width=0.8\linewidth,trim={0 1.5cm 0 1.5cm},clip]{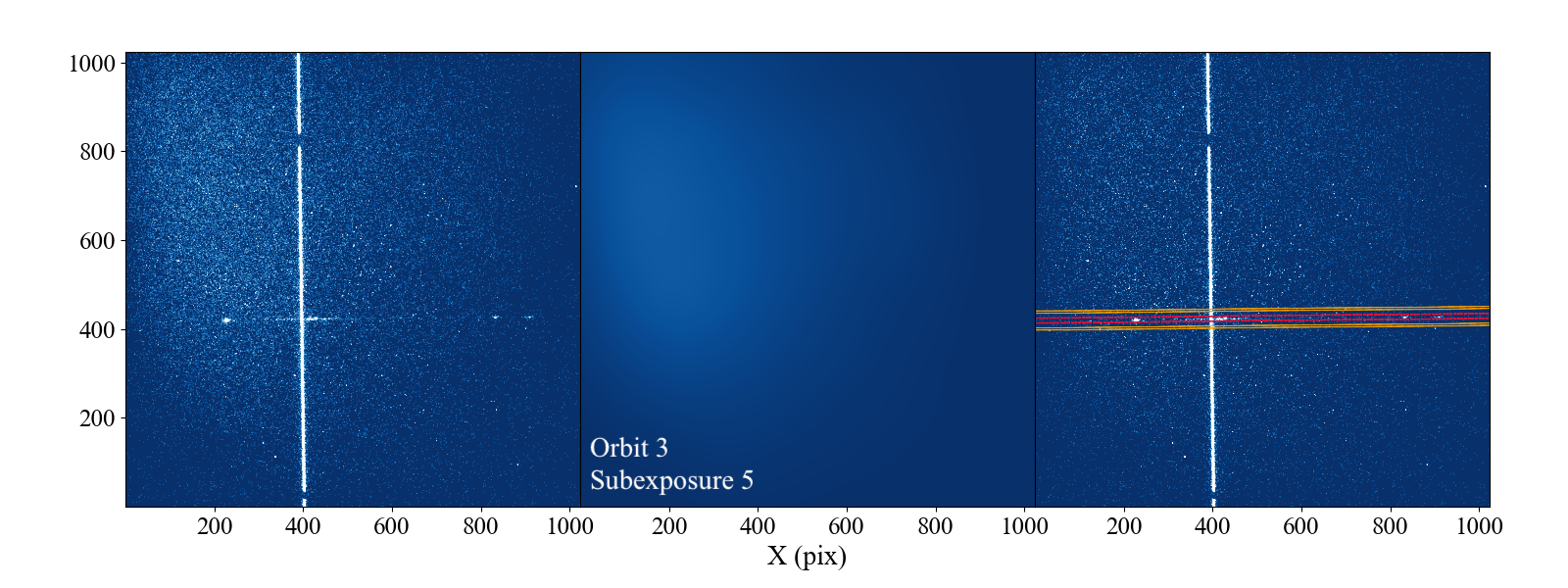}
    \includegraphics[width=0.8\linewidth,trim={0 1.5cm 0 1.5cm},clip]{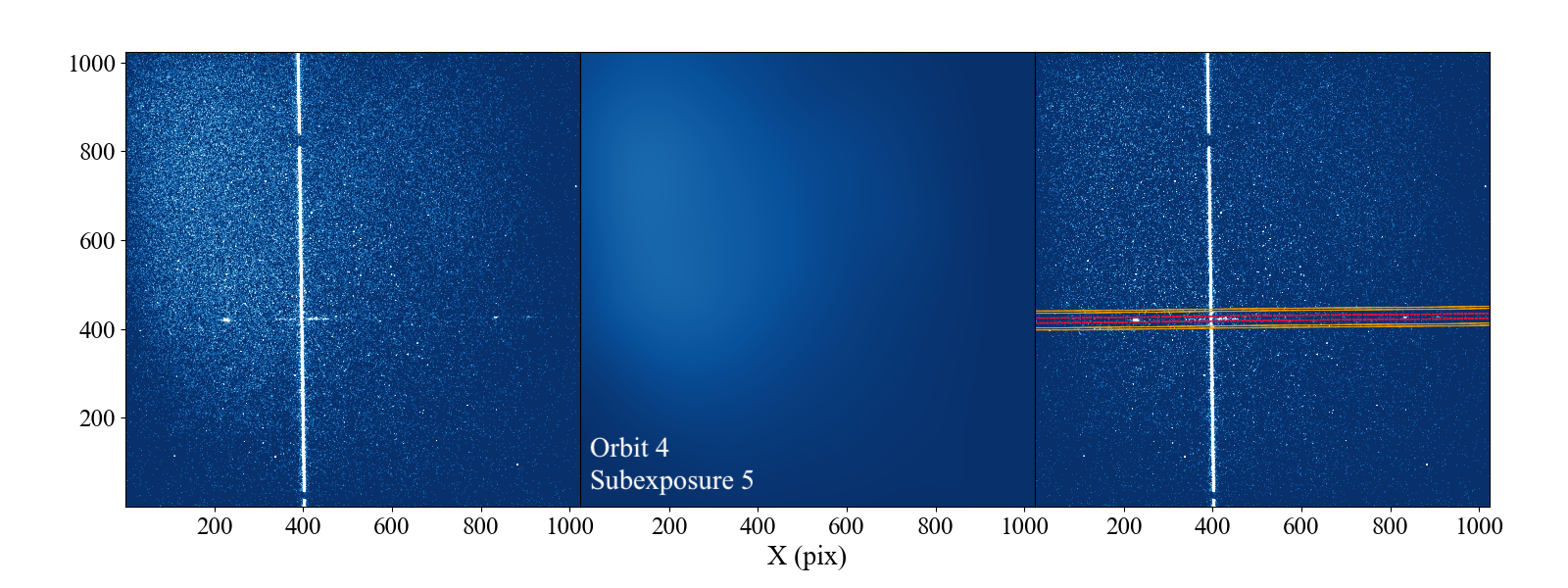}
    \includegraphics[width=0.8\linewidth,trim={0 0 0 1.5cm},clip]{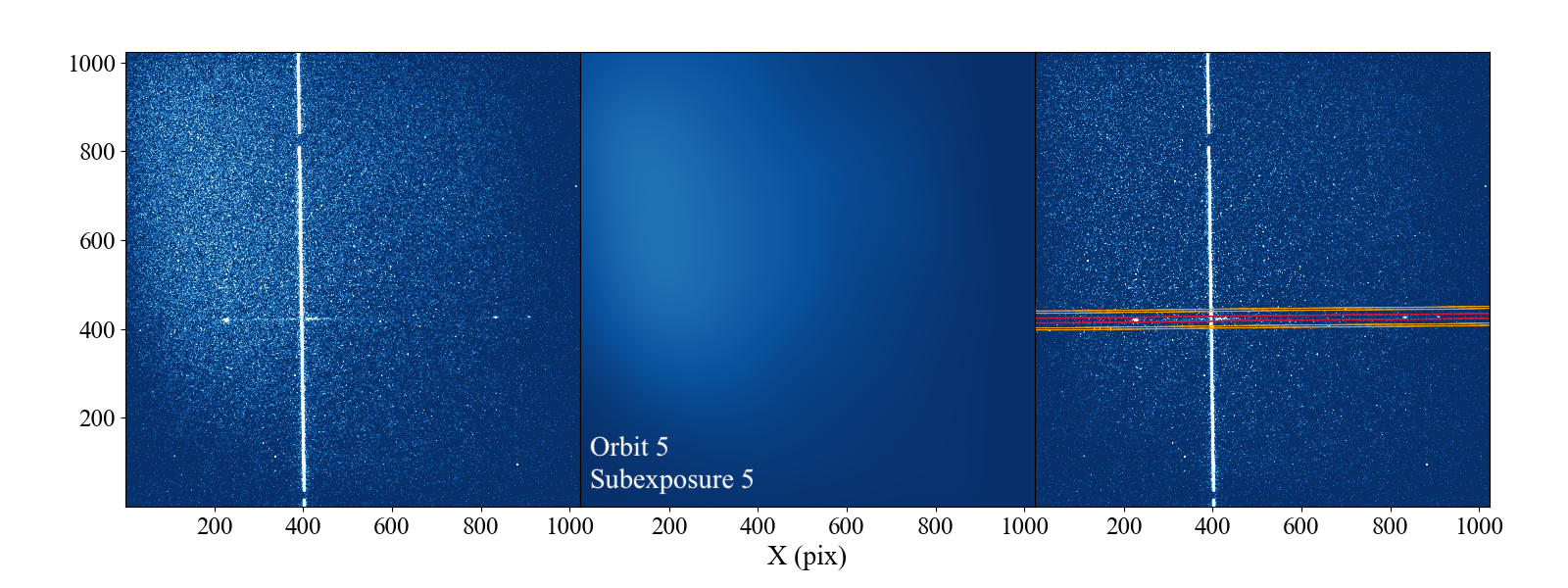}
    \caption{The \textsc{calstis} pipeline-calibrated frames exhibit significant dark-current induced background variation at the per-orbit level (left column). Individual example frames from Visit 2 is shown. Each row presents one sub-exposure (sub-exposure no.4 ie. the final sub-exposure) from each orbit of the visit. The columns show the \textsc{calstis} calibrated frame (left), the RbF model background (middle), and the final background-subtracted frame (right). Visit 1 also exhibited the same background variations, which was corrected for in the same manner.}\label{fig:background_subtraction}
\end{figure*}

To correct for this variable dark current background, we performed additional background analyses on the pipeline-calibrated individual exposures (reduced 2-D `FLT' frames). To compute a smooth background model,  the detector was sectioned into 50x50 pixel blocks, with mean values computed for each block while excluding regions affected by the vertical geocoronal emission and the horizontal spectral extraction region. Utilising radial basis function (RbF) interpolation, a model background was generated for each subexposure of each orbit based on the consistent noise profile observed across both visits. This background model was subsequently subtracted from the original data to mitigate background effects. The dark current induced background variation, and our model and corrected frames, are presented in Figure~\ref{fig:background_subtraction}.

Spectral extraction was performed with the \textsc{x1dcorr} routine applied to the newly background-calibrated observations. Background flux was estimated with the \textsc{BACKCORR} routine, which utilised a pair of background strips, 5-pixels wide and positioned $\pm$ 20\,pixels from the target trace. Multiple extraction configurations were investigated for background subtraction, with each test region producing similar light curves to the final chosen configuration. In particular, we note that the same lightcurve can be reproduced with x1d background subtraction turned off, which demonstrates that our 2-D model is sufficient in removing the background.

\subsection{\textit{HST} breathing effects}
\begin{figure}
  \centering
  \includegraphics[width=0.5\textwidth]{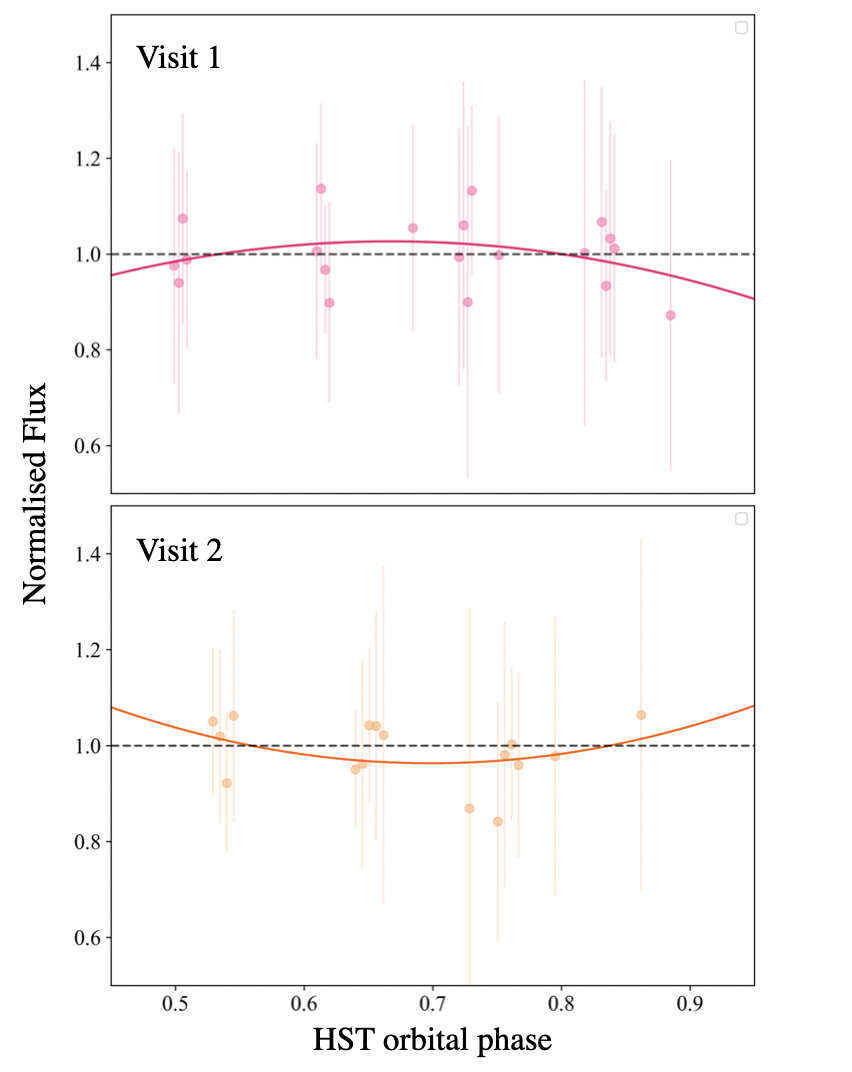}
  \caption{Lightcurves for Visits 1 and 2 phased on \textit{HST}'s orbital period to search for temporal thermal fluctuations over the course of \textit{HST}'s orbit. The solid lines represent the best fit model describing the fluctuations for each visit. The model is subsequently removed from each visit prior to light curve modelling.} 
  \label{fig:breathing}
\end{figure}

We investigated the \textit{HST} breathing effect to search for thermal fluctuations caused by \textit{HST}'s orbit \citep[e.g.][]{Kimble_1998, Bourrier_2013_b, Ehrenreich_2015, Lavie_2017, Garcia_2020, Zhang_2022, Rockcliffe_2023}. These fluctuations are induced by thermal variability in the telescope's optics, impacting the throughput of the detectors on the timescales of \textit{HST}'s orbit (95.47 minutes). To understand this variability, we took the fluxes of the sub-exposure spectra integrated over the Ly-$\alpha$ line wings, excluding the core influenced by ISM absorption. We model the overall Ly-$\alpha$ flux variability, phasefolded with the \textit{HST} orbit, with a second order polynomial for each visit, which we subsequently take into account when analysing the data further. The variations and our model correction is shown in Figure~\ref{fig:breathing}.\\

\subsection{Light curves}
\begin{figure*}[t]
  \centering
  \includegraphics[width=0.9\textwidth]{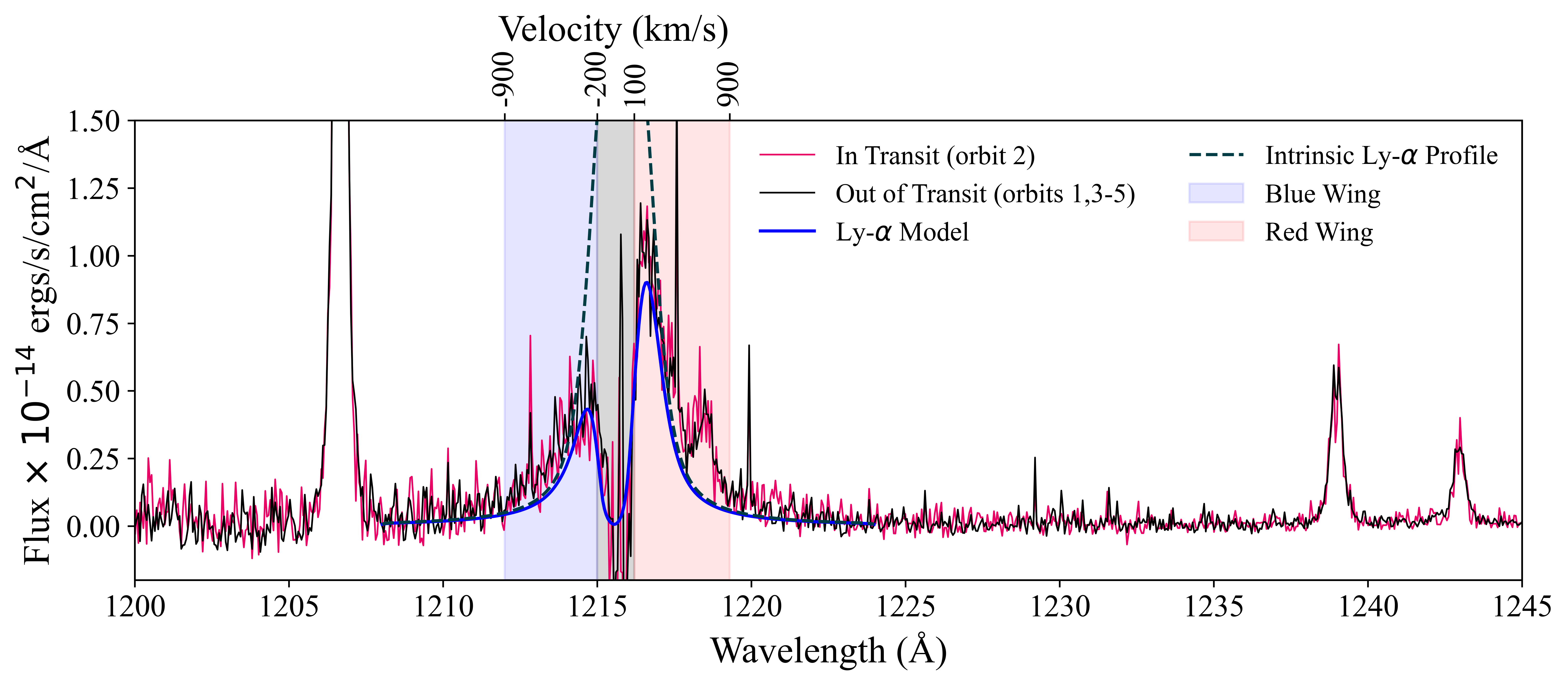}
  \caption{HIP94235 b's full FUV-MAMA spectrum is depicted in the above figure. The out-of-transit spectrum, derived from orbits 1, 3, 4, and 5 from both visits, is shown in black, while the in-transit spectrum from orbit 2 is plotted in pink, using data from both Visit 1 and Visit 2. The Ly-$\alpha$ model and intrinsic Ly-$\alpha$ profile, generated using the \textsc{lyapy} package \citep{2016ApJ...824..101Y}, are plotted in blue and navy (dashed), respectively. The shaded regions represent the blue and red wings, with corresponding velocity regions.}
  \label{fig:spectrum}
\end{figure*}

\begin{figure*}
    \centering 
    \includegraphics[width=0.83\linewidth,trim={0 0 0 0},clip]{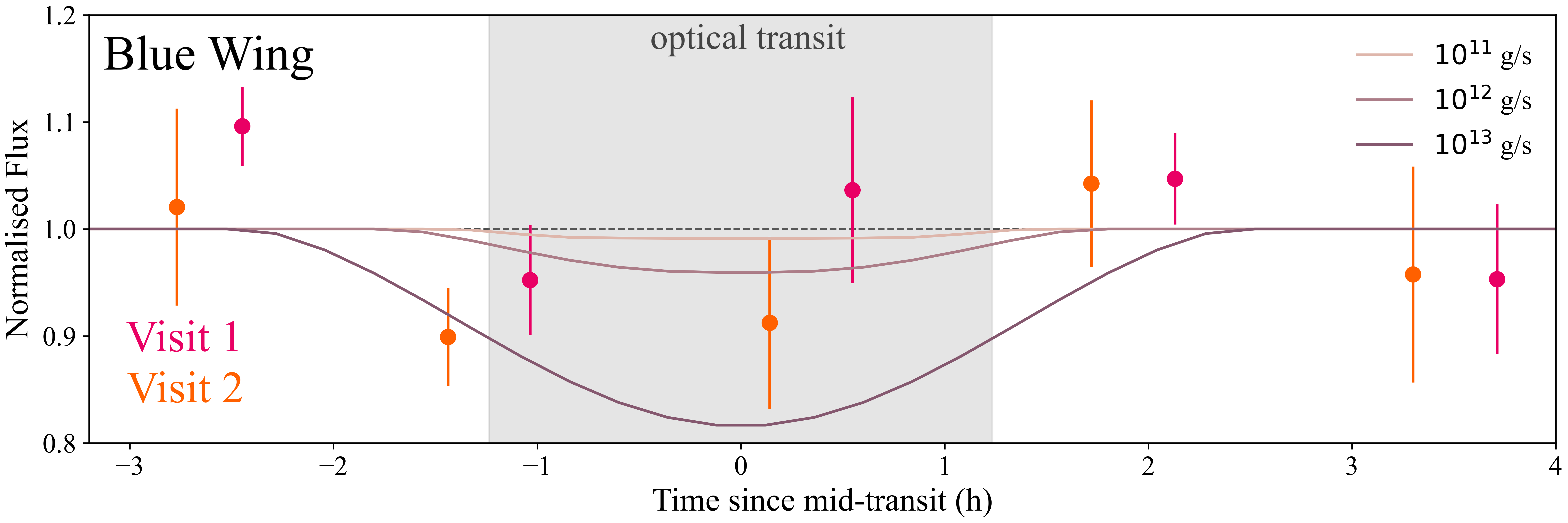}    \includegraphics[width=0.455\linewidth,trim={0 0 0 0},clip]{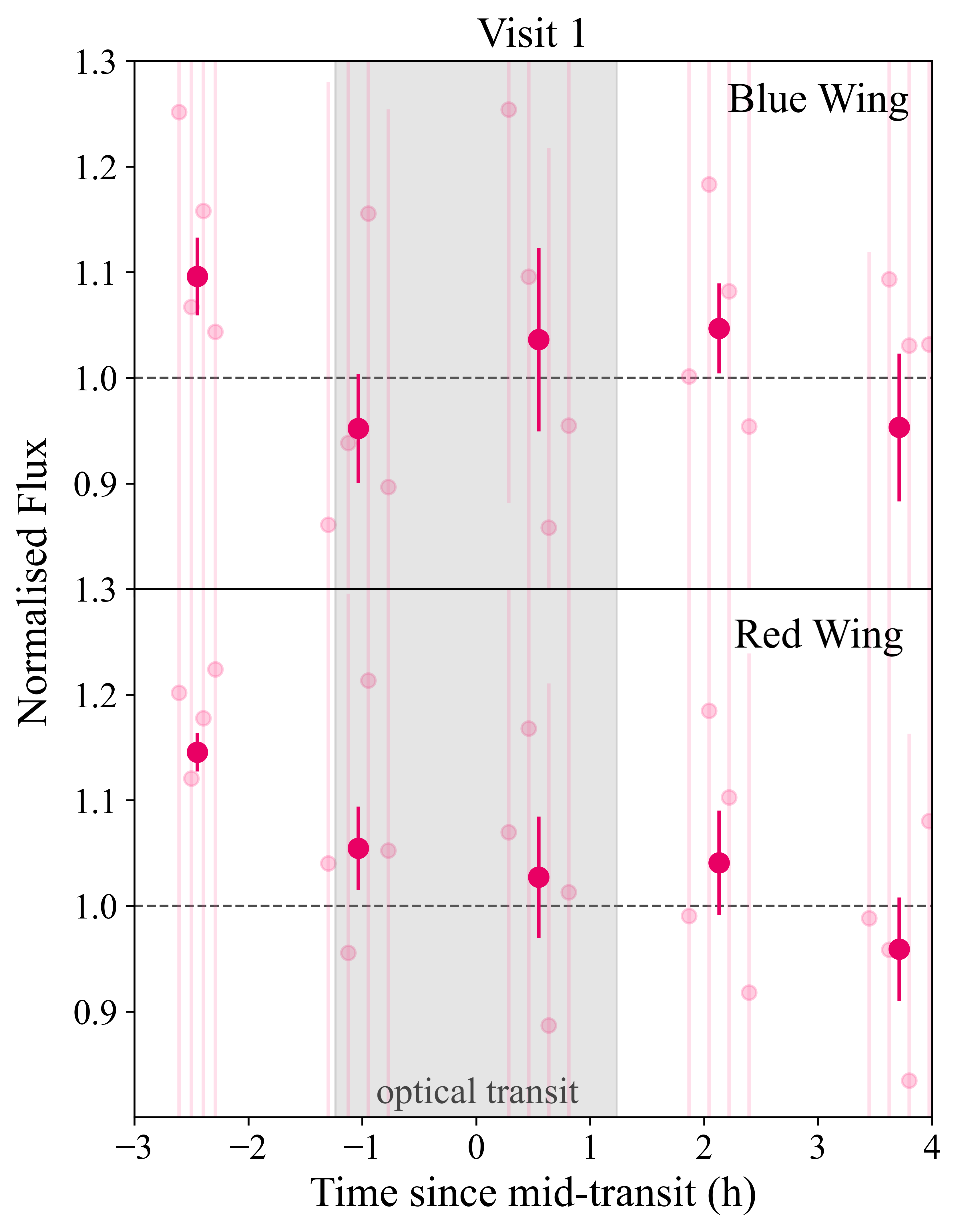}    \includegraphics[width=0.41\linewidth,trim={0 0 0 0},clip]{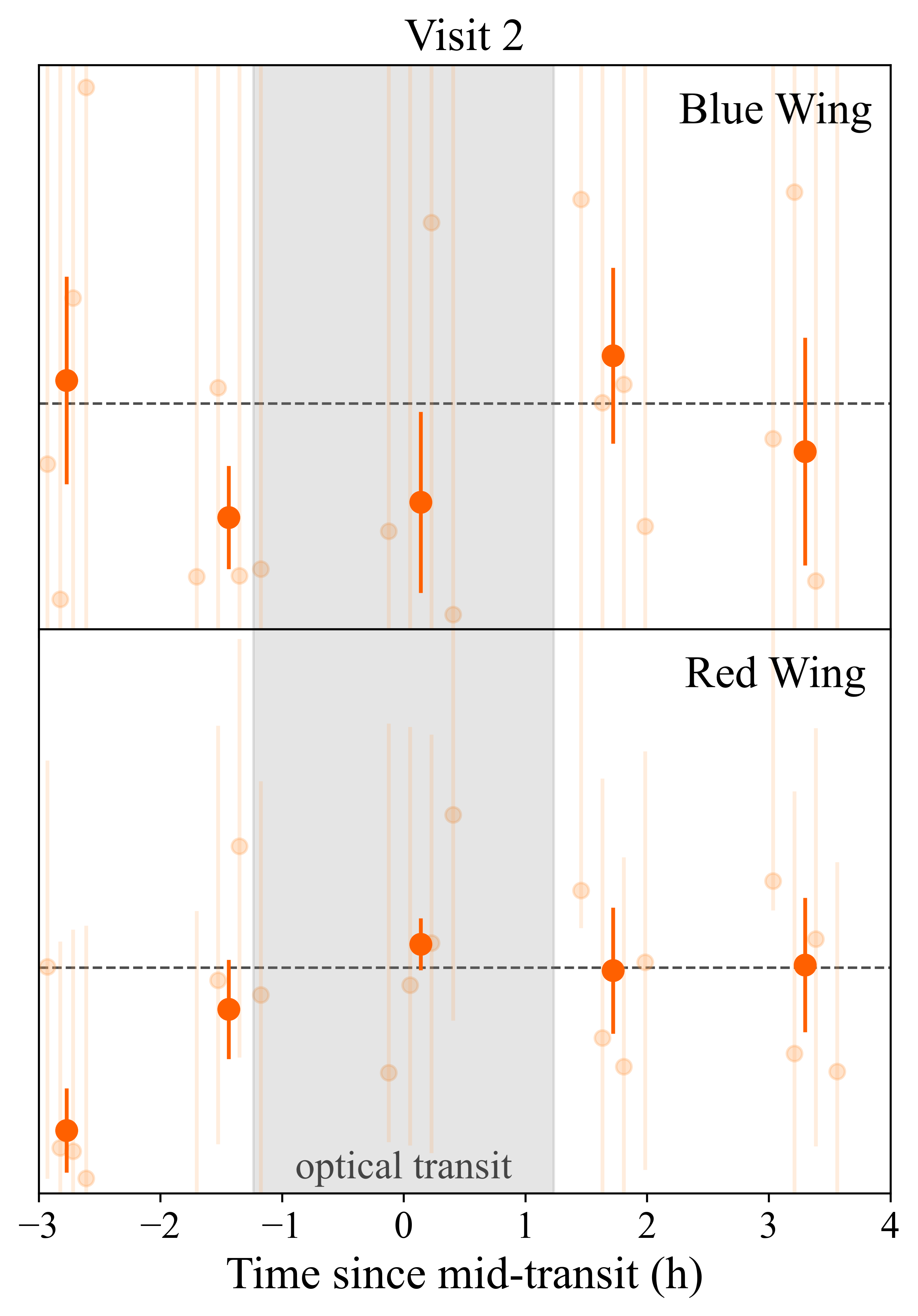}
    \caption{Transit lightcurves for the blue-wing and red-wing regions of HIP94235 b's profile, with blue and red regions defined as per Figure~\ref{fig:spectrum}. Visit 1 is shown on the left in pink, Visit 2 on the right in orange. The subexposure fluxes for each orbit are overplotted as translucent datapoints, the per-orbit bins as opaque points, each with associated errors. The top row depicts the blue wing lightcurve for both visits, with simulated transit models based off of different $\dot{M}$ models overplotted. All lightcurves are normalised to the mean of orbits 4 and 5 (the planned out of transit orbits) main points. The 2.4hr optical transit times of HIP94235 b are depicted by the shaded grey regions.}
    \label{fig:lightcurves}
\end{figure*}

We make use of the extracted 1D spectra to compute Ly-$\alpha$ light curves over each wing of the emission line. In addition, we also compute silicon III (Si III) and nitrogen V (N V) light curves to search for stellar contamination within the \textit{HST} observations. In the case of the Ly-$\alpha$ lightcurves, to ensure consistency across all sub-exposures within each orbit, we first interpolated each spectrum onto a linear wavelength grid spanning 1190-1250 \AA{}. Subsequently, the blue and red arm light curves were extracted from spectral regions of 1212-1215 \AA{} and 1216.2-1219.3 \AA{}, respectively. These wavelength regions were defined to correspond to velocity regions -200 to -900 kms$^{-1}$ and 100 to 900 kms$^{-1}$ for the blue and red wings, respectively, and were held constant over both visits. The defined spectral apertures were selected to exclude the regions affected by the interstellar absorption and geocoronal emission features located in between the wings. The blue and red wings, from which we derive the integrated light curves, are marked in Figure~\ref{fig:spectrum}. By summing the fluxes over these masked wavelength regions, we constructed the resulting lightcurves that are depicted in Fig \ref{fig:lightcurves}. We note that the lightcurve remains constant to within uncertainties despite slightly changing masked wavelength regions to test our chosen regions. In particular, we tested band passes as narrow as 1214.4-1214.9 \AA{} and 1216.0-1216.5 \AA{}, corresponding to $100\,\mathrm{km\,s}^{-1}$ width bands as close to the line core as allowed by the interstellar absorption. These narrow band passes also yielded null detections consistent with the wider band pass tests.

We find no statistically significant detection of any neutral hydrogen outflow from HIP94235 b. Our analysis reveals no discernible differences between the out-of-transit (orbits 1, 3, 4, and 5) and in-transit (orbit 2) Ly-$\alpha$ spectral line profile, as shown by Figure~\ref{fig:spectrum}. Upon integrating over the Ly-$\alpha$ wings to generate the light curves, both visits exhibit similar behaviours in the integrated blue-wing light curves, as depicted in Figure~\ref{fig:lightcurves}. We derive a 3$\sigma$ upper limit of $\sim$ 15 percent, which we use as a basis for further analysis (see Section~\ref{sec:discussion}) to constrain upper limits on mass-loss rates. Assuming similar noise characteristics, we find that one more transit would result in a 3$\sigma$ upper limit of 5 percent.

\begin{figure}[]
  \centering
  \includegraphics[width=0.4\textwidth]{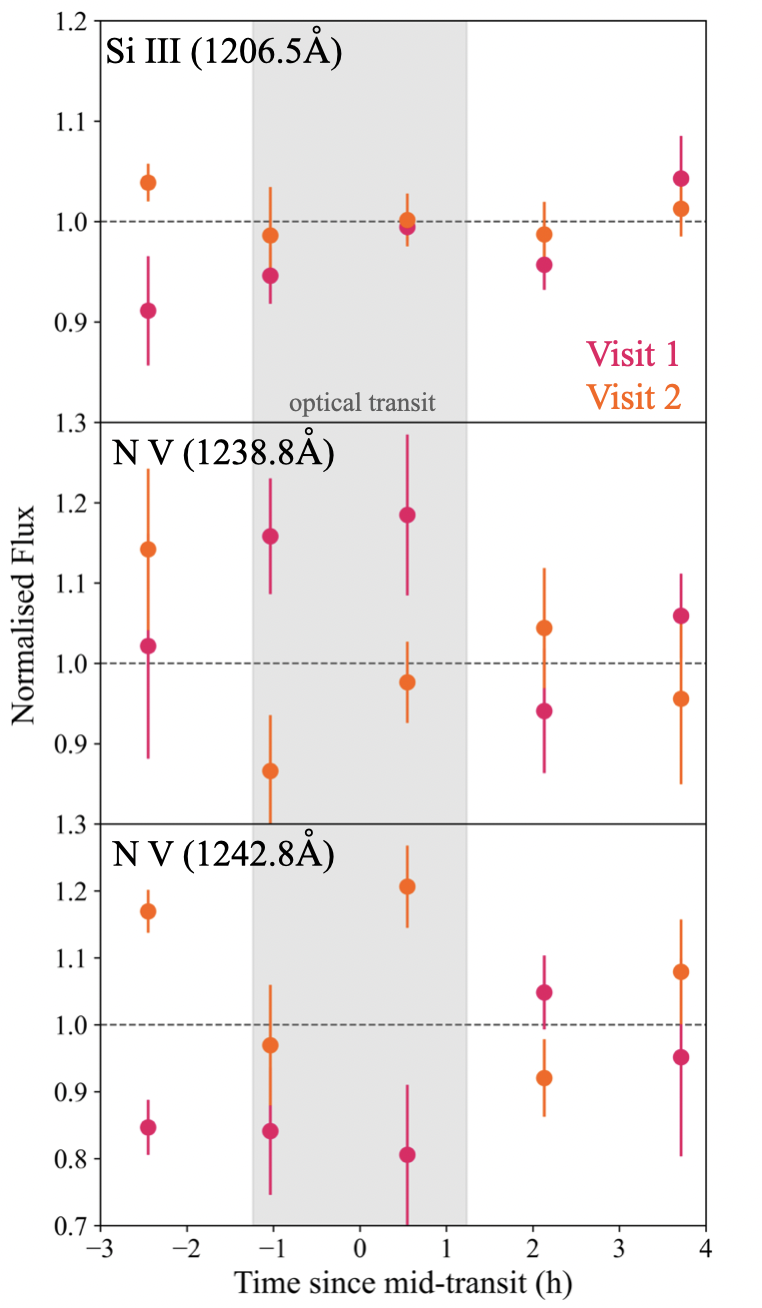}
  \caption{Transit lightcurves for the Si III(1206.3-1206.7$\AA$), N V(1238.4-1239.5$\AA$) and N V(1242.4-1243.6$\AA$) regions of HIP94235 b's profile. All three lines are visible in Fig~\ref{fig:spectrum} at their respective wavelengths. Visit 1 data is represented by the pink datapoints, with Visit 2 by the orange. Similar to before, the optical transit of HIP 94325b is depicted by the shaded grey region.}
  \label{fig:emission}
\end{figure}

Additionally, we search for excess absorption within the Si III (1206.3-1206.7 $\AA$)and N V (1238.4-1239.5 $\AA$, 1242.4-1243.6 $\AA$) lines. Traditionally, Si III and N V are regarded as sensitive tracers for stellar activity \citep{Ben-Jaffel_2013, Loyd_2014, dosSantos_2019, Bourrier_2020_MOVES}, and in recent years programs have searched for the presence of doubly ionized Si in the exosphere of exoplanets as a form of tracing hydrodynamic escape, such as in the case of HD209458 b \citep{Linsky_2010} and GJ436 b \citep{Kulow_2014, Ehrenreich_2015, Lavie_2017}. We find no variability in Si III, and no coherent variability that could be mimicking a trend in the Ly-$\alpha$ data within the N V lines, which can be seen by the band integrated lightcurves in Figure~\ref{fig:emission}. This lack of Si III and lack of coherent N V variability across both visits suggests minimal stellar activity within the period of the visits. 

\subsection{Modelling the stellar Ly-$\alpha$ line}
We model the intrinsic Ly-$\alpha$ emission profile of HIP94235 over each visit to test for any months long chromospheric variability in the host star. We make use of the \textsc{lyapy}\footnote{https://github.com/allisony/lyapy} package used in \citet{2016ApJ...824..101Y} to model the combined effects of the intrinsic line profile and the interstellar absorption jointly. 

Initially, we derived the out-of-transit quiescent line profile modeled with free parameters describing its amplitude $A$, width of the Lorentzian emission line component FWHM$_L$, the width of the Gaussian line component FWHM$_G$, velocity offset $V_n$, a self-absorption parameter $p$, and a single ISM component of fitted H I column density $\log_{10}\,N_{H I}$, width $b$, and ISM velocity shift $V_{H I}$. The ISM D I line is held constant in the modeling with a D/H ratio of $1.5\times10^{-5}$, consistent with values found by \citet{Linksy_1995}. 

The model and resulting posterior are explored via the MCMC package \emph{emcee} \citep{2013PASP..125..306F}. The best fit line profiles are illustrated in Figure~\ref{fig:spectrum}. We find an intrinsic Ly-$\alpha$ line profile of equivalent width $5.2_{-0.8}^{+1.1}\,\times10^{-14}$\AA{} on 2023-02-18, and $6.5_{-1.2}^{+1.8}\,\times10^{-14}$\AA{} on 2023-08-15. We find no variations between visits at the $1\sigma$ level.

\subsection{Updated TESS ephemeris}
\label{sec:tess}

We refine the transit ephemeris of HIP 94235 b to ensure our \textit{HST} observations fully captured the planetary transit, and that the in and out of transit Ly-$\alpha$ fluxes can be compared without ambiguity. Subsequent to the discovery paper, HIP 94235 was observed by \emph{TESS} in Sector 67 of its sixth year of operations. HIP 94235 received 20\,s cadence target pixel stamp observations during Sector 67, spanning 2023-07-01 to 2023-07-28. We make use of extracted target pixel stamp Simple Aperture Photometry (SAP; \citealt{2012PASP..124..985S,2014PASP..126..100S,2012PASP..124.1000S}) light curves processed by the NASA Science Processing Operations Center pipeline (SPOC; \citealt{jenkins2016}). 

We modeled all available \emph{TESS} and \emph{CHEOPS} light curves of HIP 94235 from Sectors 27 and 67 as per \citet{Zhou_2022}. In addition to the global model introduced in the discovery paper, we also introduced the times of each transit as free parameters, so to search for transit timing variations. 

We find a best-fit linear ephemeris of $(2037.8738\pm0.0035) + n\times (7.7130611\pm0.0000071)$. We find no evidence for transit time variations over the three year baseline. We also note the transit ephemeris from the discovery paper and that updated with new \emph{TESS} observations differ by 5 minutes at the time of \textit{HST} visit 1, consistent with the discovery paper reported uncertainties \citep{Zhou_2022}. 
 
\begin{figure}
    \centering
    \includegraphics[width=0.9\linewidth]{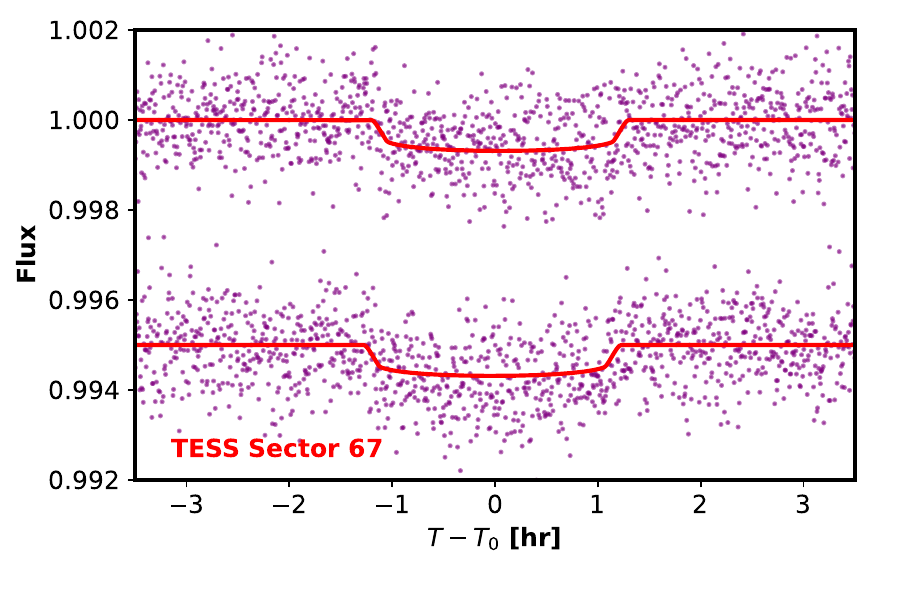} 
    \includegraphics[width=0.9\linewidth]{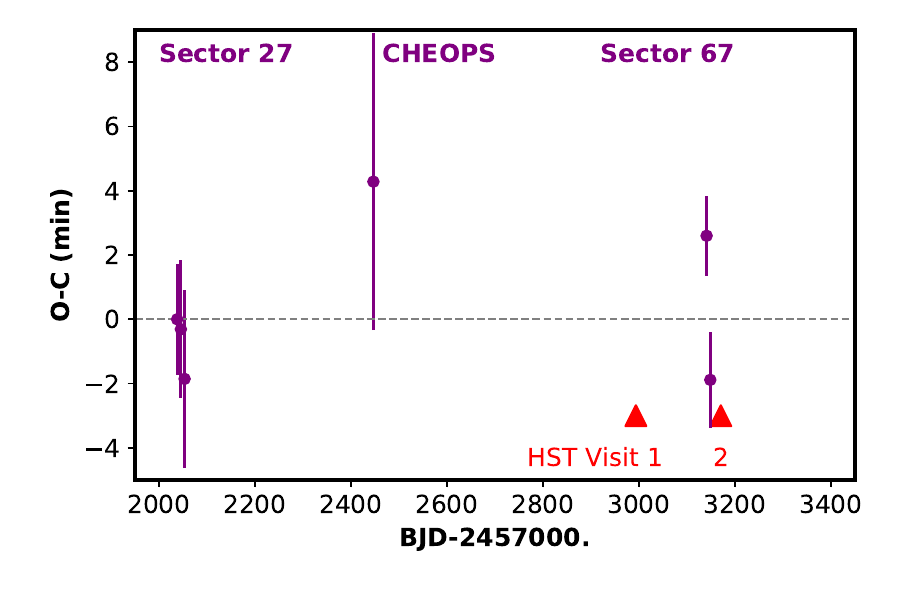} 
    \caption{Update \emph{TESS} observations and transit times for the optical transit of HIP94235 b. \textbf{Left} \emph{TESS} Sector-67 transits of HIP94235 b, observed at 20\,s cadence, and modeled as per \citet{2022AJ....163..289Z}. \textbf{Right} Individual transit times from \emph{TESS} and \emph{CHEOPS}, showing a lack of transit time deviations from a linear ephemeris. The HST visit dates are marked by the red arrows for reference.}
    \label{fig:tess}
\end{figure}

\begin{table}
    \caption{Updated transit times and ephemeris}
    \centering
    \begin{tabular}{rlll}
    \hline\hline
    \textbf{Epoch} & \textbf{Transit time} & \textbf{Uncertainty} & \textbf{Facility}\\
    \hline
    0 & 2037.8708 & 0.0012 & Sector 27\\
    1 & 2045.5837 & 0.0015 & Sector 27\\
    2 & 2053.2956 & 0.0019 & Sector 27\\
    53 & 2446.6656 & 0.0032 & CHEOPS\\
    143 & 3140.83905 & 0.00087 & Sector 67\\
    144 & 3148.5489 & 0.0010 & Sector 67\\\hline
    \multicolumn{4}{l}{\textbf{Linear ephemeris}}\\ 
    \multicolumn{4}{l}{$(2037.8738\pm0.0035) + n\times (7.7130611\pm0.0000071)$}\\
    \hline
    \end{tabular}
    \label{tab:transittimes}
\end{table}

\section{Summary and Discussion}
\label{sec:discussion}
\subsection{Mass Loss Estimates from 1-D Models based on STIS Observations}
\label{sec:pwind}
We report a non-detection of the Ly$\alpha$ transit, with no significant difference between the in-transit and out-of-transit spectra of HIP9235b (see Figure~\ref{fig:spectrum} for reference). We place a 3$\sigma$ upper limit on the transit depth at $\sim$15\%. This upper limit was derived by fitting a transit model, based on the orbital parameters found in the original discovery paper of HIP94235b by \citet{Zhou_2022}, to the blue arm data points from both STIS observations (seen in the top row of Fig~\ref{fig:lightcurves}). The radius ratio $R_p/R_{\star}$ was varied to display different planetary radii based on different mass loss rates. We find that a transit depth of 15\% would be detectable at the 3$\sigma$ level, with a corresponding mass loss rate of 10$^{13}$gs$^{-1}$.

To estimate an inferred mass-loss rate upper limit, we make use of the \textsc{p-winds}\footnote{https://github.com/ladsantos/p-winds} package \citep{2022A&A...659A..62D}. This package provides approximations of a mass-loss rate based on the planetary Ly-$\alpha$ transit depth, alongside planet and stellar parameters. It is dependant on the incident XUV flux ($0-911$\,\AA{}), which determines the hydrogen ionization fraction and the associated density profile of the outflow.

Since HIP92435 does not have its entire XUV spectrum characterised, we make use of archival observations of the 100 million year old Sun-like star EK Dra \citep{2005ApJ...622..680R} to estimate the incident XUV flux received by HIP94235 b. \citet{2005ApJ...622..680R} mapped the XUV flux of EK Dra via a combination of ASCA, ROSAT, EUVE, FUSE, and IUE observations. We calculate an incident XUV flux of $80,000\,\mathrm{erg s}^{-1}\mathrm{cm}^{-2}$ at the orbital distance of HIP94235 b, scaled from that measured for EK Dra. To test the validity of this approximation, we note that ROSAT observations of HIP 94235 \citep{Boller_2016} measure an X-ray flux of $190\pm60\,\mathrm{erg s}^{-1}\mathrm{cm}^{-2}$ (normalized at 1\, AU), fully consistent with that measured for EK Dra of $180\,\mathrm{erg s}^{-1}\mathrm{cm}^{-2}$.

Although EK Dra serves as a good approximation for HIP94235 due to its age and stellar type, uncertainties on the inferred mass loss rate remain by not knowing HIP94235's XUV flux. The X-ray flux from the youngest most active stars can be 2-3 times higher than that of EK Dra \citep[][and references therein]{2005ApJ...622..680R}. Therefore, we use this XUV flux as a lower limit when testing the effects of XUV flux on our p-winds estimate. An increase in XUV flux causes the outflow to become ionized more quickly, which in turn leads to shallower Ly-$\alpha$ transits. We assume the atmosphere is made up of only H and He, with the H number fraction set at 0.9. The temperature of the atmosphere parameter is varied from 6000-10,000 K and found to not have an impact on the resulting mass loss rate, only the width of the transit. We find a mass loss rate of $10^{13}$gs$^{-1}$ is necessary to reproduce a 15$\%$ dip in the blue-arm of HIP94235 b's Ly-$\alpha$ lightcurve. This estimate corresponds to a much higher mass loss rate than that of previous young planets. Au Mic b, HD 63433 b and HD 63433 c all have inferred mass loss rates on the order of $10^{10}$gs$^{-1}$. The transit lightcurves simulated from \textsc{p-winds} models for different mass loss values can be seen overplotted in the upper row of Fig~\ref{fig:lightcurves}. 

Fundamentally, the highly energetic XUV environment surrounding HIP94235 b rapidly ionizes any neutral hydrogen escaping from the planet, complicating the detection of substantial transits in Ly-$\alpha$. This phenomenon is hypothesised by \citet{Zhang_2022} and \citet{Rockcliffe_2021} for the non-detections of Ly-$\alpha$ outflow in HD63433 b and K2-25b, respectively. The ionization rate around HIP 94235b is the fastest of any previously surveyed planet. Following methods described by \citet{Rockcliffe_2023,refId0} for AU Mic b, and utilising our assumed high energy spectrum, we calculate a photoionization rate, $\Gamma_{ion}$, of 1.07$\times 10^{-3}$s$^{-1}$, resulting in a neutral hydrogen lifetime of 0.259 hours, or approximately 15 minutes, at the top of the atmosphere. Comparing HIP94235 b's neutral hydrogen lifetime with that of AU Mic b, which is nearly 42 minutes, highlights the impact of higher photoionization rates on Ly-$\alpha$ transit depths, which are contingent on the abundance of neutral hydrogen atoms in the outflow. The absence of a stable detection of Ly-$\alpha$ outflow from HIP94235 b, despite likely undergoing hydrodynamic escape, is thought to be attributed to its exceptionally short neutral hydrogen lifetime.

As seen through the example of photoionization above, interactions between planetary outflow and the stellar environment complicate Ly-$\alpha$ detectability. Influences such as planetary magnetic fields \citep{Owen_Adams_2014, Khodachenko_2015, Arakcheev_2017, Carolan_2021} and the ram pressure of the stellar wind \citep{McCann_2019, Khodachenko_2019, Debrecht_2020, Carolan_2021} can potentially influence the neutral hydrogen outflow. Hence, fully understanding the complex nature of Ly-$\alpha$ outflows with 1-D models is challenging, and results should be thought of as order of magnitude approximations, and as a framework to base further studies. More thorough outflow models detailing the relationship between Ly-$\alpha$ outflow and the circumstellar environment are useful in matching observational parameters to theoretical models \citep{Schreyer_2024}, while remaining computationally inexpensive. 

\begin{figure}
  \centering
  \includegraphics[width=0.5\textwidth]{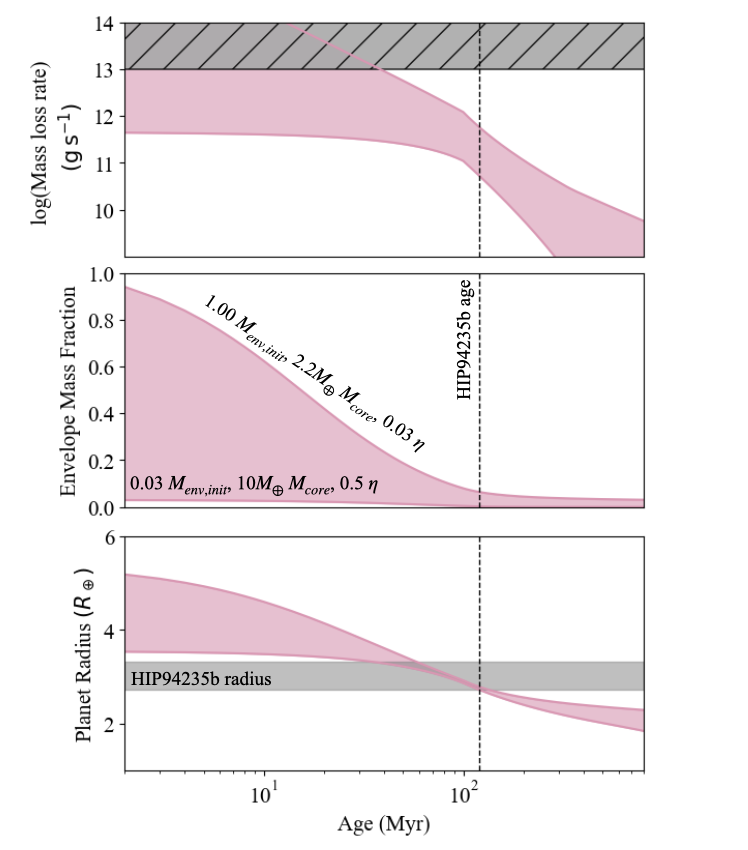}
  \caption{A range of evolution tracks for HIP94235 b. The tracks show scenarios assuming different initial envelope mass fractions ($M_{env,init}$), core masses ($M_{core}$) and mass loss efficiencies ($\eta$). Scenarios assuming gaseous envelopes result in predicted mass-loss rates of $10^{11}-10^{13}$gs$^{-1}$ at 120 Myr for HIP94235 b, similar to that inferred from p-winds (region below $10^{13}$gs$^{-1}$ in the top row plot). The hatched dark grey region above represents mass-loss values above the inferred upper limit. The range of parameters necessary to reproduce the two extreme conditions are shown.}
  \label{fig:owenwu}
\end{figure}

\subsection{Mass Loss Predictions from Energy-Limited Models}
We compare our observed mass loss rate derived from \textsc{p-winds} with predictions from energy-limited models. Figure~\ref{fig:owenwu} shows the predicted evolution of the planet radius and mass loss rate based on the energy limited analytic framework presented in \citet{Owen_2017}. Each example evolution track represents a solution for a given core mass $M_\mathrm{core}$ between $1-10\,M_\oplus$, mass-loss efficiency $\epsilon$ between 0.0-1.0 and initial envelope mass fraction $M_{env,init}$ between 0.01-1.0. 

The energy limited photoevaporation models converge with a predicted mass-loss rate between $10^{11-12} \,\mathrm{g\,s}^{-1}$ at 120 Myr (HIP94235 b's current age estimate), and are consistent with our simple \textsc{p-winds} derived upper limits allowed by our \textit{HST} observations, represented by the region below the hatched region above $10^{13}$gs$^{-1}$ in Figure~\ref{fig:owenwu}. In each case, the evolution track begins with an extensive hydrogen helium envelope, with an initial radius of $\sim 3.6-5.2\,R_\oplus$. These tracks predict that much of the initial envelope should be evaporated, with a present-day envelope mass fraction of $<10\%$. A range of core masses can reproduce the current observables, although a higher core mass requires a lower initial envelope mass and a higher eta parameter to reproduce the planetary radius observed today, as expected. For conservative models with a low atmospheric stripping efficiency ($\epsilon$ = 0.1) and a low initial envelope mass fraction ($M_{frac,env}$ = 0.1), a core mass of 5$M_\mathrm{core}$ is required to reproduce HIP94235 b's observed radius at 120Myr. Over a 1 Gyr timescale, the planet is expected to lose most of its envelope mass, traversing the radius valley and transitioning into the super-Earth regime.

\acknowledgements  
We respectfully acknowledge the traditional custodians of all lands throughout Australia, and recognise their continued cultural and spiritual connection to the land, waterways, cosmos, and community. We pay our deepest respects to all Elders, ancestors and descendants of the Giabal, Jarowair, and Kambuwal nations, upon whose lands this research was conducted.
This research is based on observations made with the NASA/ESA Hubble Space Telescope obtained from the Space Telescope Science Institute, which is operated by the Association of Universities for Research in Astronomy, Inc., under NASA contract NAS 5–26555. These observations are associated with program 17152.
AM and GZ thank the support of the ARC DECRA program DE210101893 and Future fellowship program FT230100517.
CH thanks the support of the ARC DECRA program DE200101840.

\facility{HST,TESS}
\software{emcee \citep{2013PASP..125..306F}, batman \citep{2015PASP..127.1161K}, lyapy \citep{2016ApJ...824..101Y}, p-winds \citep{2022A&A...659A..62D}, astropy \citep{2018AJ....156..123A}, PyAstronomy \citep{pya}}


\bibliographystyle{apj}
\bibliography{refs}

\end{document}